\documentclass[%
 reprint,
 amsmath,amssymb,
 aps,
pra,
]{revtex4-2}

\usepackage{graphicx}
\usepackage{dcolumn}
\usepackage{bm}

\begin{document}

\preprint{APS/123-QED}

\title{Fitness and Overfitness:\\Implicit Regularization in Evolutionary Dynamics}

\author{Hagai Rappeport}
 \email{Hagai.Rappeport@mail.huji.ac.il}
\affiliation{School of Computer Science and Engineering, The Hebrew University of Jerusalem}

\author{Mor Nitzan}
 \email{Mor.Nitzan@mail.huji.ac.il}
\affiliation{Racah Institute of Physics, The Hebrew University of Jerusalem, Jerusalem, Israe}
\affiliation{School of Computer Science and Engineering, The Hebrew University of Jerusalem}
\affiliation{Faculty of Medicine, The Hebrew University of Jerusalem, Jerusalem, Israel}
\date{\today}%

\begin{abstract}

A common underlying assumption in evolutionary thought is that adaptation generally drives an increase in biological complexity. However, the rules governing the evolution of complexity appear more nuanced, and the forces which drive increase or decrease in organismal complexity are not fully understood. 
Evolution is deeply connected to learning, where complexity, as well as its origins and consequences, are much better understood, with established results on the optimal complexity appropriate for a given learning task in various settings.
In this work, we suggest a mathematical framework for studying the relationship between organismal complexity and the complexity of the environment in which they evolve by leveraging an existing mathematical isomorphism between evolutionary dynamics and learning theory, specifically, an isomorphism between the replicator equation and sequential Bayesian learning, with evolving types corresponding to competing hypotheses and fitness in a given environment to likelihood of observed evidence.
In Bayesian learning, implicit regularization prevents overfitting and drives the inference of hypotheses whose complexity matches the learning challenge. We show how these results naturally carry over to the evolutionary setting, where they are interpreted as organism complexity evolving to match the complexity of the environment, with organisms which are too complex or too simple for their environment suffering from effects we term \textit{overfitness} and \textit{underfitness}, respectively. Other aspects, peculiar to evolution and not to learning, reveal additional trends. One such trend is that frequently changing environments decrease selected complexity, a result with potential implications to both evolution and learning. 
Together, our results suggest that the balance between complex organisms that may over-adapt to transient environmental features, and simple organisms that may be insufficiently flexible in
responding to environmental challenges, drives the emergence of optimal complexity which reflects
environmental structure. This mathematical framework offers new ways of thinking about biological complexity, and suggests new potential causes for it to increase or decrease in different selective environments.

\end{abstract}

\maketitle

\section{\label{sec:intro}Introduction \protect}

Living systems, across scales and levels of organization, are complex. This complexity, manifested in morphological, organizational and behavioral aspects of organisms, is hailed by many as one of life's defining features~\cite{kauffman_origins_2011}. But while our detailed understanding of various complex adaptations continually expands, a complete theory of the nature and origins of biological complexity itself remains elusive~\cite{mcshea_complexity_1991}. As with virtually any other feature of living systems, biological complexity may make more sense in light of its evolution \cite{wolf2018physical, szathmary1995major, adami2000evolution}.

Early thinkers on evolution assumed that lineages tend to progressively increase in complexity as they adapt to environmental challenges~\cite{darwin1859, spencer_principles_2020, mcshea_complexity_1991}, and much work has been dedicated to elucidate the evolutionary forces behind this supposed trend ~\cite{lenski_evolutionary_2003, carroll_chance_2001}. Biological reality is however, as always, more nuanced. Parasites, simpler in many aspects than their free-living relatives~\cite{keeling_simplicity_2004}, are a standard example of environments favoring reduced complexity. In general, however, we do not fully understand the selective pressures affecting the increase or decrease of complexity and their interplay in various contexts and environments~\cite{mcshea_complexity_1991, veit2025evolution}.

The existing literature on selection for decreased complexity has tended to focus on two points. The first is a supposed maintenance cost of complexity and the potential fitness advantages gained from forfeiting it and diverting resources elsewhere~\cite{wagner_energy_2005, lynch_bioenergetic_2024}. The second point relates to the entropic force generated on genomes by the constant barrage of mutations which accumulate unless purged by stabilizing selection. Presumably, any complex structure not under strong stabilizing selection is thus expected to erode over time~\cite{muller_genetic_1932, moran_accelerated_1996, maughan2007roles}. While these two evolutionary forces are certainly relevant, it remains unclear what their relative contribution is and whether they are the only ones at play~\cite{heylighen_growth_1999}. Complexity may also theoretically hinder adaptation rates due to pleiotropy \cite{orr_adaptation_2000}, but empirically this seems not be the case \cite{wagner2008pleiotropic} (see also \cite{lenski_evolutionary_2003}) . 

An additional hurdle to developing a theory of biological complexity is a lack of an agreed-upon definition of the subject matter. Multiple formalisms have been suggested, but none have met universal acceptance. Biological complexity can be defined as the amount of information a genome carries about its environment \cite{adami_what_2002}. But under such a definition complexity necessarily increases until genomes encode a maximal amount of information on their environment, thus not capturing the phenomenon of potentially decreasing complexity in simple environments. One path to a standalone definition of organismal complexity has been to transfer to a biological context the idea, well understood in theoretical computer science, of complexity as a measure of a system's minimal description \cite{hinegardner1983biological, grunwald_minimum_2007, simon_architecture_1962,dingle_inputoutput_2018}. Other definitions appeal to a notion of complexity as related to the "number of parts" of which an organism is composed of \cite{godfrey-smith_complexity_1998}, be they genes \cite{bird1995gene, brem2005landscape, van2019biological}, cell types \cite{carroll_chance_2001}, developmental pathways \cite{mcshea1996perspective} or morphological traits \cite{valentine1994morphological}. 

The drawback of all aforementioned definitions is that they are external to evolutionary theory, and consequentially their causal interaction with selection is unclear. In particular, they do not allow for making predictions of evolutionary trends in complexity depending on different types of environments. To bridge this gap, a notion of complexity which is directly emergent from evolutionary dynamics is required.

The “evolutionary dynamics” in question can be captured by many different models, but arguably the simplest~\cite{nowak_evolutionary_2006} is the discrete-time replicator equation:
\begin{equation}
x_{i}^{\left(t+1\right)}\propto x_{i}^{\left(t\right)}f_{i}^{\left(t\right)}   \label{eqn:replicator}
\end{equation}

\noindent with $ x_{i}^{\left(t\right)}$ representing the frequency of type $i$ at generation $t$ and $f_{i}^{\left(t\right)}$ the fitness of type $i$ given the environment experienced at that generation. The form of this dynamical equation is analogous (and indeed mathematically equivalent) to a different system where types change in frequency over time in response to external constraints favoring some over others. Namely, in the context of Bayesian inference, we write
\begin{equation}
p\left(H_{i}\mid E\right)\propto p\left(H_{i}\right)p\left(E\mid H_{i}\right)  \label{eqn:bayes}
\end{equation}
\noindent where $p\left(H_{i}\mid E\right)$ is the posterior of hypothesis $i$ given its prior $p\left(H_{i}\right)$ and the likelihood $p\left(E\mid H_{i}\right)$ that $H_i$ assigned to the observed evidence $E$. Sequential inference is then described by the update equation 

\begin{equation}
p^{\left(t+1\right)}\left(H_{i}\right)\propto p^{\left(t\right)}\left(H_{i}\right)p\left(E^{\left(t\right)}\mid H_{i}\right)  \label{eqn:bayes_seq}
\end{equation}

\noindent where  $p^{\left(t\right)}\left(H_{i}\right)$ represents the probability associated with hypothesis $i$ at timestep $t$ and $p\left(E^{\left(t\right)}\mid H_{i}\right)$ the likelihood under hypothesis $i$ of the evidence observed at that timestep.

Eq. \ref{eqn:replicator} and Eq. \ref{eqn:bayes_seq} describe the same exact dynamics, only with a different interpretation of the quantities being updated \cite{harper_replicator_2010}. However, facts about the dynamics which may be obvious in one scope may be completely obscure in the other and vice versa. In particular, Bayesian learning has a well developed theory of optimal complexity for a given learning task. Over the past decade a rich literature has emerged on connections between evolution and learning theory, porting insights from the latter to better understand the former~\cite{watson_how_2016, kouvaris_how_2017, czegel_bayes_2022, campbell_universal_2016, watson2014evolution}. Here we will attempt to expand this intellectual enterprise to the study of the evolution of complexity. 

Learning theory has multiple operational notions of complexity~\cite{akaike_new_1974, burnham_model_2004, vapnik_nature_1995, grunwald_minimum_2007}, relating a model's (hypothesis') complexity to the number of degrees of freedom it possesses. A common practice in computational learning is to penalize, or \textit{regularize}, models to prevent them from becoming too complex, as models which are complex relative to the underlying task are likely to \textit{overfit}. That is, they may fit noise in the training data rather than the underlying patterns~\cite{bishop_pattern_2006}. In the analogy to evolution, such penalties correspond to direct fitness costs of maintaining complex structures, as discussed above. In addition to explicit regularization, many forms of learning also possess implicit forms of regularization emergent from their learning dynamics ~\cite{bishop_pattern_2006, mackay_information_2003}. The question naturally arises, whether there exist analogs of implicit regularization in evolutionary dynamics, and if so, what is their significance to the evolution of complexity?

\section{\label{sec:model}Model \protect}
Denoting type $i$'s phenotype at generation $t$ as $y_i{\left( t \right)}$, we assume fitness decreases monotonically with distance to an optimal phenotype $y^*{\left( t \right)}$, which is therefore the unique fitness peak for that generation. Our goal is to follow how selective pressures vary for organisms of different complexity depending on the complexity of their environment. To this end, we define organismal and environmental complexity as the number of degrees of freedom in the processes generating organism phenotypes $y_i{\left( t \right)}$ and optimal phenotypes $y^*{\left( t \right)}$, respectively. Measuring model complexity via the number of free parameters required to specify it is standard in learning theory \cite{akaike_new_1974, burnham_model_2004} and closely corresponds to algorithmic complexity \cite{grunwald_minimum_2007, wolpert_rigorous_1992}. We defer further treatment of this measure and its place in the literature on biological complexity to the discussion section.  

The most straightforward way to model environments and genotypes in this setting is as generative distributions over phenotype space. However, phenotype distributions provide only a statistical summary of outputs without relating to the underlying processes that generate these responses. To better capture the mechanistic relationship between sensing, processing, and phenotypic expression, a more natural setting is populations of \textit{functions}. This aligns naturally with computational learning, where one typically evaluates competing functions on their ability to map inputs to outputs rather than just examining their output distributions.

\begin{figure*}[t!]
\centering
\includegraphics[width=0.9\linewidth]{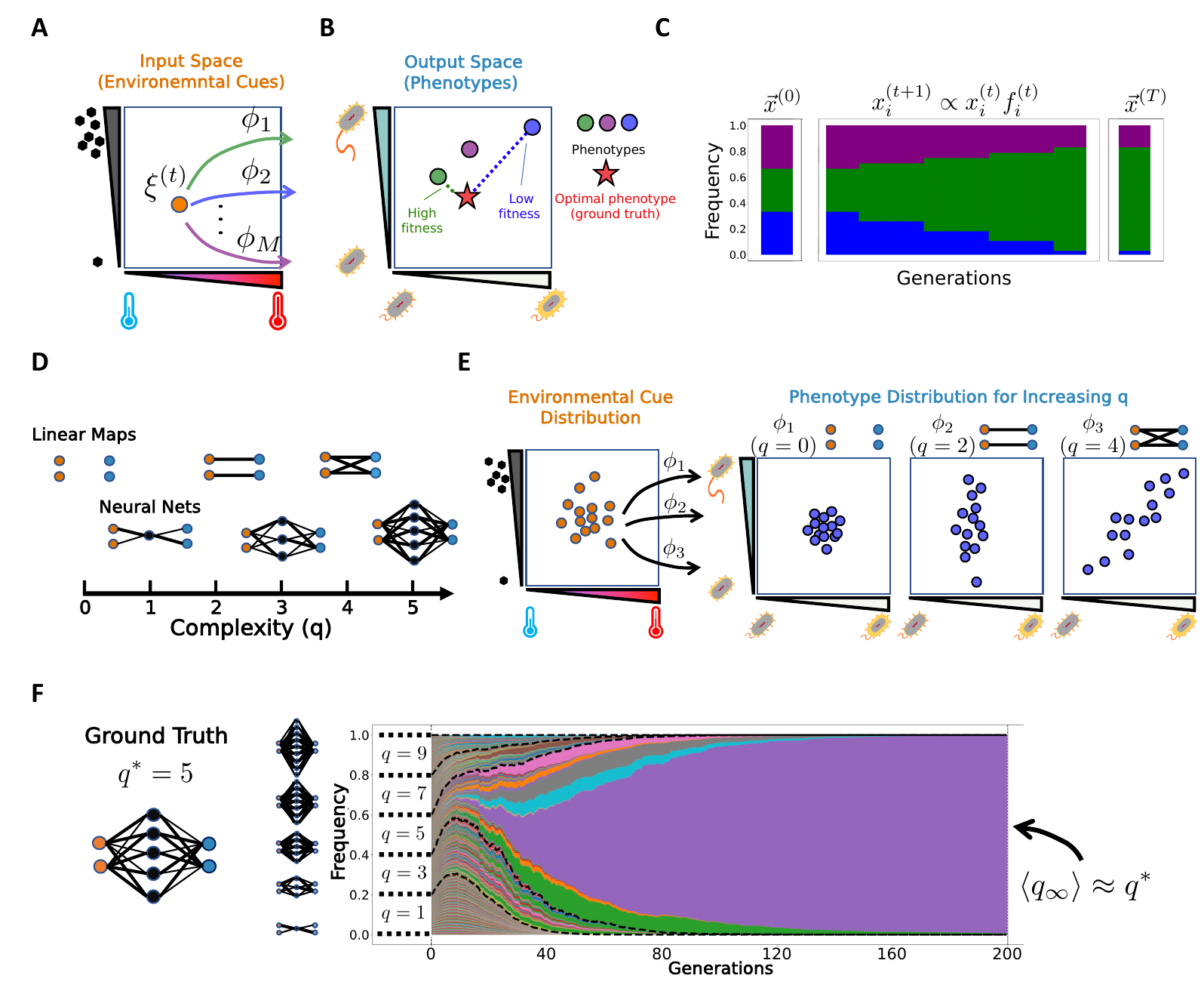}
\caption{\textbf{A model for selection of complexity in environments of a given complexity}.\\
 \textbf{(A)} Populations consist of functions $\left\{ \phi_{i}\right\}$ mapping environmental cues $\xi^{\left(t\right)}$ (e.g. temperature or nutrient availability) received at each timepoint $t$ to phenotypes (e.g. membrane thickness or flagellum size). \textbf{(B)} Fitness in a given environment is defined via a ground truth function $\phi^*$ (optimal by definition) and each member’s fitness $f_i$ is an increasing function of the distance between its phenotype to the optimal phenotype $\phi^{*}\left(\xi^{\left(t\right)}\right)$. \textbf{(C)} Selection follows the discrete time replicator equation, and we follow relative frequencies of each member. Illustrated is a muller plot following the dynamics of a population composed of three members starting from a uniform distribution $\vec{x}^{(0)}$ and evolving into a population with distribution  $\vec{x}^{(T)}$. \textbf{(D)} Complexity is defined as the number of tunable parameters in a function, here illustrated for two function classes - linear mapping (top) or 1-hidden-layer neural networks (bottom). \textbf{(E)} Different members $\phi_i$ may have different associated complexities $q_i$, resulting in different levels of intricacy for their mappings from environment to phenotype space. The environmental complexity $q^*$ is defined as the complexity of $\phi^*$. \textbf{(F)} Muller plot for a typical simulation of the replicator dynamics, following a population of functions divided into 5 complexity classes (dashed lines) and starting from a uniform distribution.  The simulation culminates with the dominance of a member from the $q=5$ class, which is equal in this case to $q^*$.}\label{fig:intro}
\end{figure*}

Organisms infer features of the environment they find themselves in by various sensing apparatus, and adjust their phenotypes accordingly. Thus, we consider genotypes as coding for input-output maps $\phi_i$ (fig \ref{fig:intro}A), where inputs are environmental cues (either measurements of external conditions such as temperature, nutrient availability etc. or signals conveyed by other organisms) and outputs are phenotypes (any variant morphologies or behaviors which may possibly affect fitness). The optimal phenotype in a given environment is defined via a "ground truth" map $\phi^*$, which encodes the underlying environmental regularities, i.e. the optimal response for any possible environmental cue (fig \ref{fig:intro}B), and fitness decreases monotonically with distance to the optimal phenotype.

In this setting, we follow pure selection dynamics (with no mutation) according to the replicator equation over populations of parametrized functions $\left\{ \phi_{i}\right\} _{i=1}^{M}$ , with parameters representing genomes (fig \ref{fig:intro}C). The population is divided into $N$ \textit{complexity classes} $\left\{ C_{k}\right\} _{k=1}^{N}$, where each class is defined by the number of parameters required by its functions in   (fig \ref{fig:intro}D,E). More parameters allow for capturing more intricate patterns in mapping environmental cues into phenotypes (fig \ref{fig:intro}E), such as in e.g. the degree of intricacy required for advanced combinatorial sensing \cite{malnic1999combinatorial}. The \textit{environmental complexity} is then defined as the complexity of $\phi^*$ , matching the intuition  that the complexity of an environment should match the intricacy of response required to thrive in it. In the interest of focusing on selection, this setting does not consider \textit{generating} novel complex phenotypes \cite{orr_adaptation_2000}, an issue we leave for future work. 

At each generation $t$, all functions $\left\{ \phi_{i}\right\} _{i=1}^{M}$ receive a randomly sampled environmental cue $\vec{\xi}^{\left(t\right)}$, and the fitness of a given member $\phi_i$ is given by a gaussian selection term $f_{i}^{\left(t\right)}=e^{-\gamma\left\Vert \phi_{i}\left(\xi^{\left(t\right)}\right)-\phi^{*}\left(\xi^{\left(t\right)}\right)+\epsilon^{\left(t\right)}\right\Vert }$ \cite{xue_environment--phenotype_2019}. The prefactor $\gamma$ determines selection strength; smaller values of $\gamma$ lead to smaller fitness differences and hence smaller population changes per timestep. When generations are measured in units of $\frac{1}{\gamma}$ , the dynamics become independent of $\gamma$, in the limit $\gamma\to  0$  converging to the continuous-time stochastic replicator equation (methods). Lastly, $\epsilon^{\left(t\right)}\sim\mathcal{N}\left(0,\sigma\right)$ is a noise term, modeling features relevant for fitness which are not encoded by the provided environmental cues.

\section{\label{sec:results}Results \protect}
For simplicity sake, and to best illustrate the underlying principles, we start by considering $\left\{ \phi_{i}\right\} _{i=1}^{M}$ and $\phi^*$ to be linear mappings. We note, however, that all results in this section hold qualitatively for other types of mappings such as polynomials or feed-forward neural networks, see SI appendix for further details. In addition, in what follows we consider the setting in which all complexity classes start with the same representation in the population, as the effects in play are clearer in this setting. See the SI appendix for a demonstration of the insensitivity of the main findings to this assumption.

\subsection*{Implicit Regularization in the Replicator Equation Prevents Overfitness}
\begin{figure*}
\centering
\includegraphics[width=.85\linewidth]{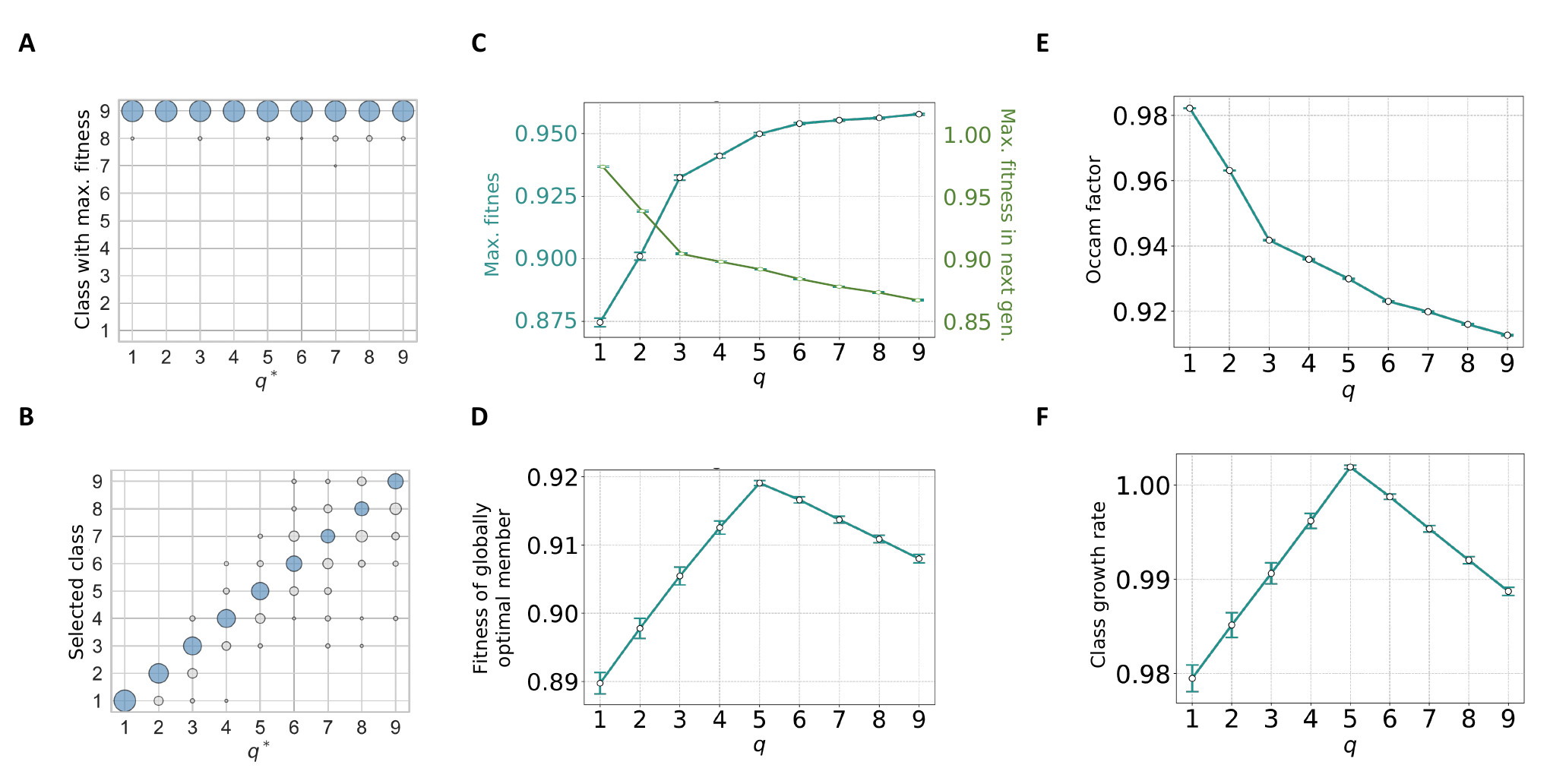}
\caption{
\textbf{Selected complexity depends both on a class' maximal per-timestep fitness as well as on fitness stability over time.}
\textbf{(A)} The maximal fitness is typically found in the highest complexity class, independent of $q^*$. Circle sizes represent, for a given $q^*$ (x-axis), the fraction out of 100 simulations in which time-averaged max fitness was found in a given complexity class (y-axis), with highest fraction marked in blue. \textbf{(B)} Selected complexity (y-axis), defined as the average population complexity at the simulation's end. The selected class tends to correspond to the environmental complexity ($q^*$; x-axis) in which selection occurred. Circle sizes and colors as in (A). \textbf{(C-F)} Illustration for a specific $q^*$ ($q^*=5$). \textbf{(C)} While maximal fitness is strictly increasing with the complexity of the class (y-axis, left), high fitness at a given timestep does not necessarily predict high fitness at the next, illustrated by plotting fitness at time $t+1$ of the member which achieved best fitness at time $t$ (divided by its fitness at time $t$) (y-axis, right). \textbf{(D)} The best globally optimal member (with highest time-averaged fitness) is found in q*. \textbf{(E)} The Occam factor (y-axis) decreases with complexity (x-axis). \textbf{(F)} Average class growth rate (which is proportional to the maximal fitness times the Occam factor) is peaked at q*, which thus eventually dominates.
Results in all panels are based on 100 simulation realizations with linear maps. Population size was 2,000 per complexity class, $\sigma=1$, $\gamma=0.05$ and $T=1,000$.  In (C-F) circles represent mean values and errorbars  95\% confidence intervals (1.96 standard errors).}
\label{fig:analysis_fixed_env}
\end{figure*}

In all function classes considered, at any given timestep the maximal fitness achieved is found in the class of highest complexity $q_\text{max}$ (fig \ref{fig:analysis_fixed_env}A. See SI appendix for a discussion on the effect of population size on maximal fitness). Nevertheless, the complexity class which eventually comes to dominate the distribution (hereinafter the \textit{selected complexity}, denoted  $\left\langle q_{\infty}\right\rangle$) is typically not $q_\text{max}$, but rather corresponds to environmental complexity, i.e. $\left\langle q_{\infty}\right\rangle \approx q^*$ (fig \ref{fig:analysis_fixed_env}B).

To understand this apparent discrepancy, we find it illustrative to start by considering, for each complexity class $C_k$, the member of $C_k$ with maximal fitness at timestep $t$, call it $i_k^*{(t)}$ , and follow its fate at the next timestep $t+1$ (for the following discussion, we focus on a fixed, intermediate, environmental complexity $q^*=5$). While the fitness  of  $i_k^*{(t)}$ indeed increases with complexity, its fitness at $t+1$ (normalized to its fitness in timestep $t$) decreases with complexity (fig \ref{fig:analysis_fixed_env}C). When considering the fitness of each class’s \textit{globally} best performing member (i.e. the one with the highest mean log fitness over the whole simulation), we indeed find the best performing member in $q^*$ (fig \ref{fig:analysis_fixed_env}D). As a whole, this analysis suggests that the apparent increased success of complex classes is only due to their ability to better fit specific environmental cues or noise realizations rather than the underlying environmental challenge.

To formalize and better understand this effect, we next turn to a mathematical construct which is closely related to the aforementioned ratio of fitness values, known in learning theory as the \textit{Occam factor} \cite{wolpert_rigorous_1992}. In terms of the replicator equation (eq. \ref{eqn:replicator}), for a setting such as ours, dividing our population into $N$ subpopulations (classes) $\left\{ C_{k}\right\} _{k=1}^{N}$, we can write a replicator equation for the frequencies of classes with the form

\begin{equation}
X_{k}^{\left(t+1\right)}\propto X_{k}^{\left(t\right)}F_{k}^{\left(t\right)}
  \label{eqn:replicator_class}
\end{equation}

\noindent where $X_{k}^{\left(t\right)}$ is the frequency of class $k$ at generation $t$ (relative to other classes) and $F_k$ serves as class fitness, i.e. the growth rate of class $C_k$ once normalized to other classes' fitness. Following the notation introduced above, the frequency of the highest-fitness member of class $C_k$ at generation $t$ is $x_{i_{k}^{*}\left(t\right)}$ and we denote $\tilde{x}_{i_{k}^{*}\left(t\right)}^{\left(t\right)}=\frac{x_{i_{k}^{*}\left(t\right)}^{\left(t\right)}}{X_{i_{k}^{*}\left(t\right)}^{\left(t\right)}}$, i.e. $\tilde{x}_{i_{k}^{*}\left(t\right)}^{\left(t\right)}$ is within-class relative frequency. We can now write (see derivation in methods section)
\begin{equation}
F_{k}^{\left(t\right)}=f_{i_{k}^{*}\left(t\right)}^{\left(t\right)} \frac{\tilde{x}_{i_{k}^{*}\left(t\right)}^{\left(t\right)}}{\tilde{x}_{i_{k}^{*}\left(t\right)}^{\left(t+1\right)}}  \label{eqn:occam}
\end{equation}

That is, the fitness of class $k$ at time $t$ is a product of the fitness of its best member at that timestep, $f_{i_{k}^{*}\left(t\right)}^{\left(t\right)} $, and a factor of magnitude less than $1$, the \textit{Occam factor}, which is the reciprocal of the change in frequency in favor of the current best member $i_{k}^{*}\left(t\right)$. A smaller Occam factor corresponds to a larger "collapse" of a class's internal distribution in favor of its current best member, and is associated with decreased class growth at that timestep. 

The fitness of members in complex classes tends to change more substantially over time compared to simple classes (fig \ref{fig:analysis_fixed_env}C), and at any given timestep $t$, complex classes typically contain members which score high in fitness at time $t$ but have scored relatively low in the past and are therefore present with low frequency (fig \ref{fig:analysis_fixed_env}C). Thus complex classes collapse on new “optimal” members, every generation anew, and overall have lower average Occam factors (fig \ref{fig:analysis_fixed_env}E). Consequently, they possess suboptimal class growth rates (fig \ref{fig:analysis_fixed_env}F), explaining the fact that increased complexity beyond environmental complexity does not lead to increased selection. In analogy to the term used in learning theory, we term this drawback of complexity “overfitness”. 

To summarize this section, we find that even without imposing any external cost on complexity, the dynamics of the replicator equation are sufficient to select, over time, complexity which matches the underlying environmental complexity. This occurs via two opposing forces. On the one hand, overfit classes which are too complex relative to environmental complexity favor a different member each timestep and therefore do not rise to dominance. On the other hand, classes which are too simple are unable to match nuances in the optimal mapping (\textit{underfit} the environmental challenge) and are therefore disfavored.

\subsection*{Simple Classes are Transiently Favored in the Initial Phase of Selection}
\begin{figure*}[t!]
\centering
\includegraphics[width=.85\linewidth]{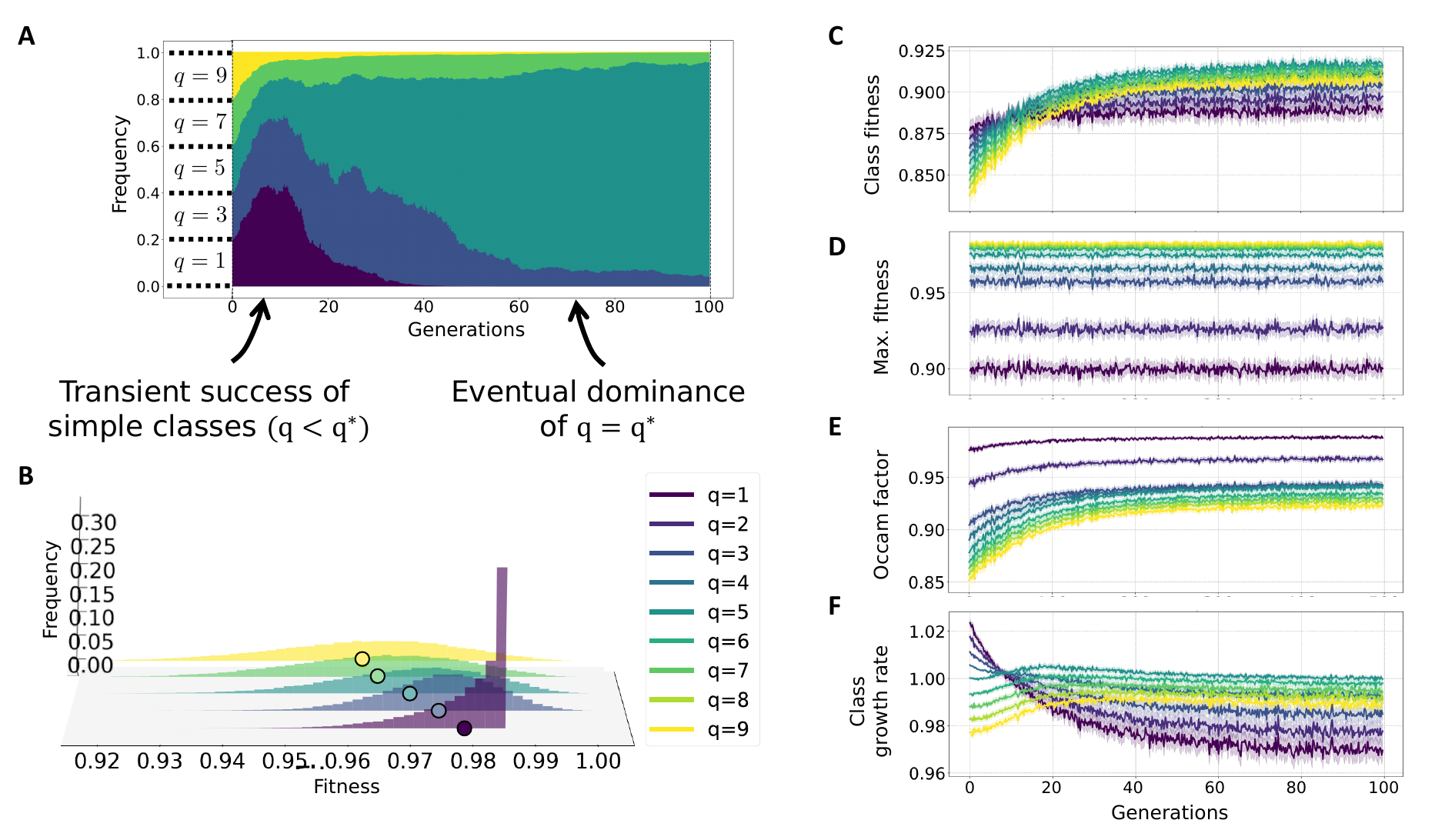}
\caption{
\textbf{Simple Classes are Transiently Favored in the Initial Phase of Selection.}
\textbf{(A)} Muller plot of class frequencies in a typical simulation realization. While the complexity of the member which eventually dominates selection is  $q^*$ (here $q^*=5$), in the initial phase of selection there is a transient success (increase in frequency) of classes with $q<q^*$. \textbf{(B)} Fitness distribution of different classes with respect to a randomly chosen ground truth function. More complex classes have a wider fitness distribution with a higher maximal fitness value, but a lower mean (colored circles). 
\textbf{(C-F)} Dynamics (averaged over realizations) of the metrics shown in fig \ref{fig:analysis_fixed_env} .
\textbf{(C)} Class fitness changes over time, initially highest for simple classes before converging to favor the class with complexity equal to $q^*$. \textbf{(D-F)}. Class fitness can be decomposed into the product of maximal class fitness \textbf{(D)} and its Occam factor \textbf{(E)}, and normalized to give a class's growth rate \textbf{(F)}.
Results in (B-F) are based on 100 simulation realizations with linear maps. Population sizes of 2,000 per complexity class, $\sigma=1$, $\gamma=0.05$ and $T=500$.   
}
\label{fig:transient}
\end{figure*}

Thus far we have considered only the long-term success of different complexity classes. In an evolutionary setting, short-term selection dynamics may prove relevant as well. In a typical simulation realization, before the $q=q^*$ complexity class eventually comes to dominate, we  observe a transient period where simple complexity classes are more successful (fig \ref{fig:transient}A). A first hint towards explaining this transient is that we may write a class's fitness (eq. \ref{eqn:replicator_class}) as the average fitness of its members, weighted by their current representation in the class:

\begin{equation}
F_{k}^{(t)}=\frac{\sum_{i\in C_{k}}x_{i}^{(t)}f_{i}^{(t)}}{X_{k}^{(t)}} \label{eqn:class_fit_1}
\end{equation}

Our simulations begin from a uniform composition for all populations and therefore at $t=0$ we have $F_{k}^{(0)}\propto \langle p_k \rangle$, i.e. the fitness of class $k$  is equal to the mean of the fitness distribution $p_k$ from which it drew its values, and this mean value is highest for simple classes (fig \ref{fig:transient}B). Thus, at $t=0$ the simplest classes exhibit the largest growth rate (fig \ref{fig:transient}F).

At $t>0$ class fitness begins to change due to selection, and via a close analog of Fisher's fundamental theorem of natural selection \cite{fisher1999genetical} (see methods for derivation)

\begin{equation}
\Delta F_{k}\left(0\right)=\frac{Var\left[p_{k}\right]}{\left\langle p_{k}\right\rangle }  \label{eqn:class_fit_2}
\end{equation}
Since complex classes have higher values for $Var\left[p_{k}\right]$ and lower values for $\left\langle p_{k}\right\rangle$ (see also in fig \ref{fig:transient}B), their rate of class fitness increase outpaces simple classes and start overtaking them.

In the long term, a class $C_k$'s fitness equals the fitness of its optimal member $i^*_k$ (methods). Since $ f_{i_k^*}$  is highest for the $q=q^*$ class  (fig \ref{fig:analysis_fixed_env}D), after sufficient time this class dominates the overall population composition, as demonstrated above. 

A different way of understanding this effect is by following the dynamics of the Occam factor. A class's growth rate is a product of the fitness of its best member and its Occam factor. While maximal fitness does not change with time (fig \ref{fig:transient}D), the Occam factor increases for all classes as their composition becomes concentrated around their respective best members (fig \ref{fig:transient}E) at which point the growth rate of all classes reaches its steady state value (fig \ref{fig:transient}E). Thus, the transient in question represents the initial process of locating optimal members for the given environment, where simple classes possess an advantage.

To summarize, selection starts with a transient phase where complex classes have yet to promote optimal members in their inner composition and consequentially simple classes are selected at their expense. This effect is similar to the one discussed in \cite{burger1995evolution, chevin2010adaptation}, where the mean phenotype of a population in a changing environment lags behind the optimal phenotype. Here, however, using language and tools of learning theory allows us to explicitly analyze the dependence of this effect on organism complexity.


\subsection*{Rapidly Changing Environments Select for Lower Complexity}
\begin{figure*}[t!]
\centering
\includegraphics[width=.85\linewidth]{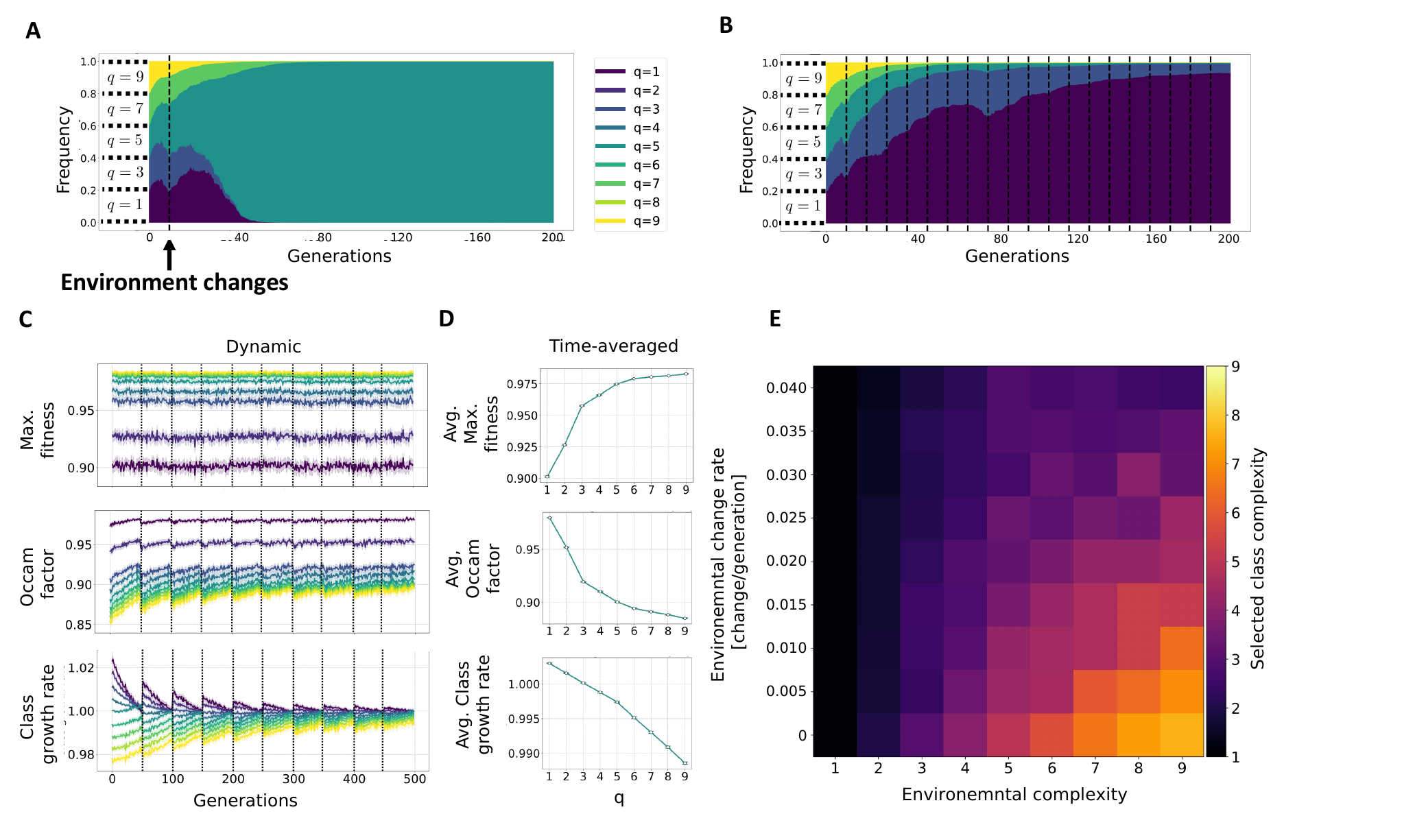}
\caption{
\textbf{Environmental Change Selects for Reduced Complexity}. 
\textbf{(A)} Muller plot  of class frequencies (as in fig \ref{fig:transient}), with the environment (the ground truth function $\phi^*$) changed to a different randomly sampled one at $t=50$. When the ground truth function changes (vertical dashed line), the transient success of simple classes repeats for a second time before subsiding. \textbf{(B)} When the environment changes at the same timescale as that of the transient, the transient becomes fixed and simple classes come to dominate selection. \textbf{(C)} Dynamics of the same metrics as in fig \ref{fig:transient}D-F, only with environments changing every 50 timesteps. \textbf{(D)} Time-average values of the metrics in (C). \textbf{(E)} Selected class (average class at the end of selection) as a function of both environmental complexity (x-axis) and environmental change rate (here, the number of different environments in a simulation of length $T=1000$) (y-axis). 
Results in (C-E) are based on 100 simulation realizations with linear maps. Population sizes of 2,000 per complexity class, $\sigma=1$, $\gamma=0.05$ and $T=500$.  
}
\label{fig:changing_env}
\end{figure*}

As detailed above, the transient success of simple complexity classes reflects a period in which optimal members of each class in the given environment do not yet dominate their respective classes. We next asked if this transient would repeat if the environment, i.e. the  ground truth function, changed mid-selection. So far, our model captured a certain degree of environmental change at each timestep, via the environmental cue which follows changes in environmental parameters. However the underlying regularities in optimal response to these parameters, i.e. the ground truth function, was hitherto assumed constant. Changing the ground truth thus models a drastic environmental change or a local shift to a different environmental niche, one that could change the fitness landscape for a given set of environmental parameters. In our model, fitness is uncorrelated in two randomly chosen environments,  so when the environment changes each class effectively “starts over” in discovering its new best member. Indeed, changing the environment in such a manner may lead to a second period of success for simple classes (fig \ref{fig:changing_env}A). However, this success is transient as well, eventually dying out and giving way to a member of the complexity class that matches the environmental complexity . 

Given this effect, the natural next question was what would happen if environments kept changing repeatedly, faster than the rate at which the transient dies out. Under these conditions, the transient success of simple classes is no longer transient, and the system may reach an equilibrium dominated by classes with lower complexity than environmental complexity (fig \ref{fig:changing_env}B). 

Maximal fitness per timestep monotonically increases with class complexity, and is unaffected when the environment changes, provided the new environment is of the same complexity as the previous one (fig \ref{fig:changing_env}C, top). However, a new environment implies new optimal members, and consequentially a drop in the Occam factor (fig \ref{fig:changing_env}C, middle) until said optimal members rise in frequency. Since this effect is stronger in complex classes, when the environment changes sufficiently fast (relative to the timescale of the transient) the Occam factor of the complex classes never surpasses that of the simple classes, and the latter's growth rate remains highest (fig \ref{fig:changing_env}C, bottom).

Quantifying this effect globally, by measuring the selected class for each value of $q^*$ and for a range of environmental change rates, we find for all values of  $q^*$ a trend of decreased selected complexity with increasing environmental change rate (fig \ref{fig:changing_env}E). 

To summarize, we find a consistent pattern of a transient “lag” period, where simple classes succeed at the expense of more complex ones, even if the latter eventually come to dominate. We may interpret this lag as the time it takes complex classes to locate the correct hypothesis (or the optimal member, as it were) for a newly presented environmental challenge. When environments change at a rate comparable to this lag period, complex classes may always contain the optimal member per environment, but by the time they are able to promote this member in frequency and start growth, the environment changes again. Simple classes, on the other hand, maintain internal variability and are much less affected when environments change. Thus, the internal complexity of the environment and the rate at which it changes are both relevant to selection of complexity, but have opposite effects.

\section{\label{sec:discussion}Discussion \protect}

Recently, multiple lines of work in theoretical evolution have made use of mathematical isomorphisms which formalize the analogy between evolution and learning ~\cite{watson_how_2016, kouvaris_how_2017, czegel_bayes_2022, watson2014evolution}, allowing one to port theorems and results from one field into the other. Given that in learning theory, complexity is well-defined and many of its effects understood, these isomorphisms present a hitherto untapped opportunity to transfer insights on complexity into evolutionary theory. A hypothesis' complexity in learning is defined via the number of tunable parameters it requires. We have adopted this definition, interpreting it as the number of tunable parameters in the environment-phenotype map. This definition instantiates a common theme considered in the literature on biological complexity, heterogeneity of possible states and responses to external stimuli \cite{godfrey-smith_complexity_1998}.   

Using this mathematical framework, we showed how evolutionary dynamics in their simplest form contain implicit mechanisms that prevent both \textit{overfitness} and \textit{underfitness}. Namely, organisms with too many tunable parameters will tend to tune these parameters, via evolution, to match patterns in their current environment which are no more than temporary noise, not indicative of future challenges. Simpler organisms are more robust to this effect, but may be too simple for a given environment, unable to utilize environmental information which is useful as well as enduring. The optimal organismal complexity is found balancing these two selective forces, and is dictated by the environmental complexity.

Implicit regularization of complexity is well understood in computational learning \cite{mackay_information_2003, wolpert_rigorous_1992}. However, the transient effects in the initial phase of selection we found are, to the best of our knowledge, not studied within this literature. Indeed, such transients bear little significance in learning of a single given task with sufficient training time (which is still the most common setting in computational learning, but perhaps not for long with the rise of online or continual learning challenges). In an evolutionary setting the “learning task” (e.g. the correct behavior in a given ecological niche) often changes on a timescale commensurable with the selection timescale. It is therefore natural in this setting to investigate the effect of changing environments \cite{burger1995evolution, chevin2010adaptation, mustonen_fitness_2009}, which we find to be substantial, and to interact in non-trivial ways with population complexity.

The notion of “environmental complexity” as well as “environmental change rate” are both considered important in evolutionary dynamics in general and the evolution of complexity in particular, but they are mostly considered closely related (e.g. \cite{dridi2016environmental, godfrey2013environmental}. But see \cite{adami_what_2002} for a notable exception). Our results suggest that they should be considered separately, as their effects may act in opposite directions.

There are multiple factors which are relevant to evolution of complexity but were not discussed here. We have focused on the selective value of complexity rather than its origins, and have therefore assumed that  relevant variability already preexists in the population and have not modeled mutations or recombination (but see SI appendix for a results concerning invasion scenarios). In addition, our framework does not model frequency-dependent selection. This last point may prove especially important as it is a prerequisite for arms-races, often speculated as drivers of increased complexity \cite{dawkins1979arms, heylighen_growth_1999}. We leave the incorporation of these effects for future work.

In our model, organisms could sense environmental parameters but not which environment, containing which regularities, they are currently in. In the presence of signals carrying such information, organisms which are able to sense and utilize them may be favored \cite{bergman1994evolution}. For example, a short-lived creature such as a fruit fly (\textit{Drosophila melanogaster}) may have genetic variants which are differentially adapted to different seasons, and consequentially populations may experience annual selective sweeps favoring different genomes at different seasons \cite{bergland_genomic_2014}. In contrast, more complex organisms sense seasonal change and adapt accordingly. A natural extension of our framework would be to incorporate such effects.

We believe the framework introduced here allows for tackling some aspects of complexity evolution in a more first-principles manner than was previously possible. In particular, it provides a framework with natural definitions of both organismal and environmental complexity, definitions which may be instantiated in different ways in different models, as well as translated into observables to be followed in natural and experimental systems.

\section*{\label{sec:acknowledgements}ACKNOWLEDGMENTS}
We wish to thank Nathalie Balaban and Roy Friedman for insightful discussions.

\appendix 
\section{\label{sec:other_function_classes}Implicit Regularization in Diverse Function Classes}
The results in the main text focused on selection of linear functions. However, the phenomena described are universal (qualitatively), across different types of functions. We  demonstrate this using populations of polynomials and 1-hidden-layer neural networks (see methods for implementation details).

\begin{figure}[t!]
\centering
\includegraphics[width=0.45\textwidth]{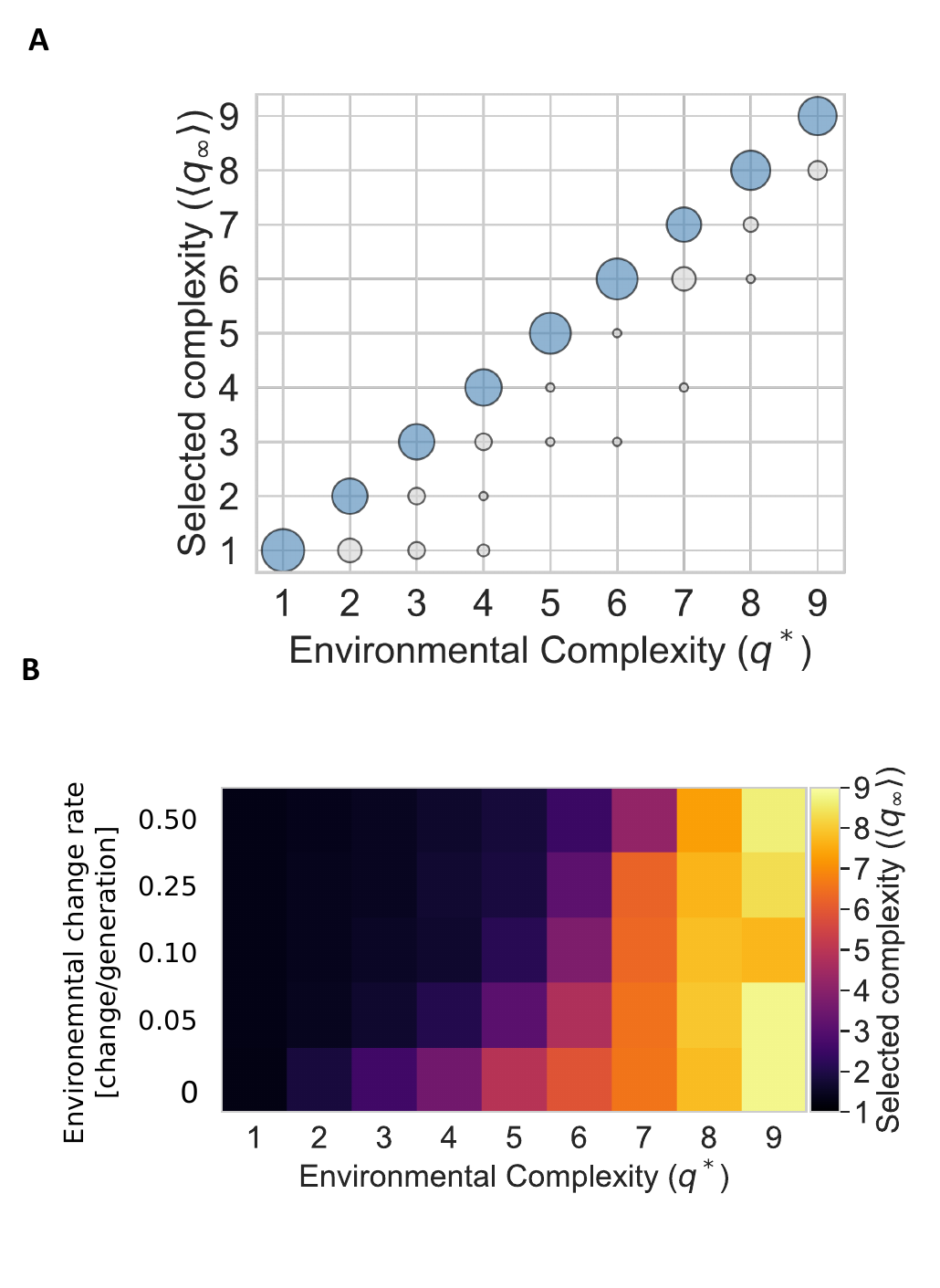}
\caption{
\textbf{Selected complexity in stable and changing environments for polynomials}. 
\textbf{(A)}  Circle sizes represent, for a given $q^*$ (x-axis), the fraction out of 100 simulations in which time-averaged max fitness was found in a given complexity class (y-axis), with highest fraction marked in blue.
\textbf{(B)} Selected class (average class at the end of selection) as a function of both environmental complexity (x-axis) and environmental change rate (here, the number of different environments in a simulation of length $T=1000$) (y-axis).
Results are based on 100 simulation realizations. Population sizes of 500 per complexity class, $\sigma=0.1$, $\gamma=0.05$ and $T=1000$.  
}
\label{fig:polynomials}

\end{figure}

Polynomials were chosen as an illustrative case as they are a classic example of the effect of overfitting \cite{hastie2009elements}. Here, we see the same behavior as with the linear functions. Namely, selected class corresponds closely to the environmental complexity (fig \ref{fig:polynomials}A) and increasing environmental change rate decreases selected complexity (fig \ref{fig:polynomials}B).

\begin{figure}[t!]
\centering
\includegraphics[width=0.45 \textwidth]{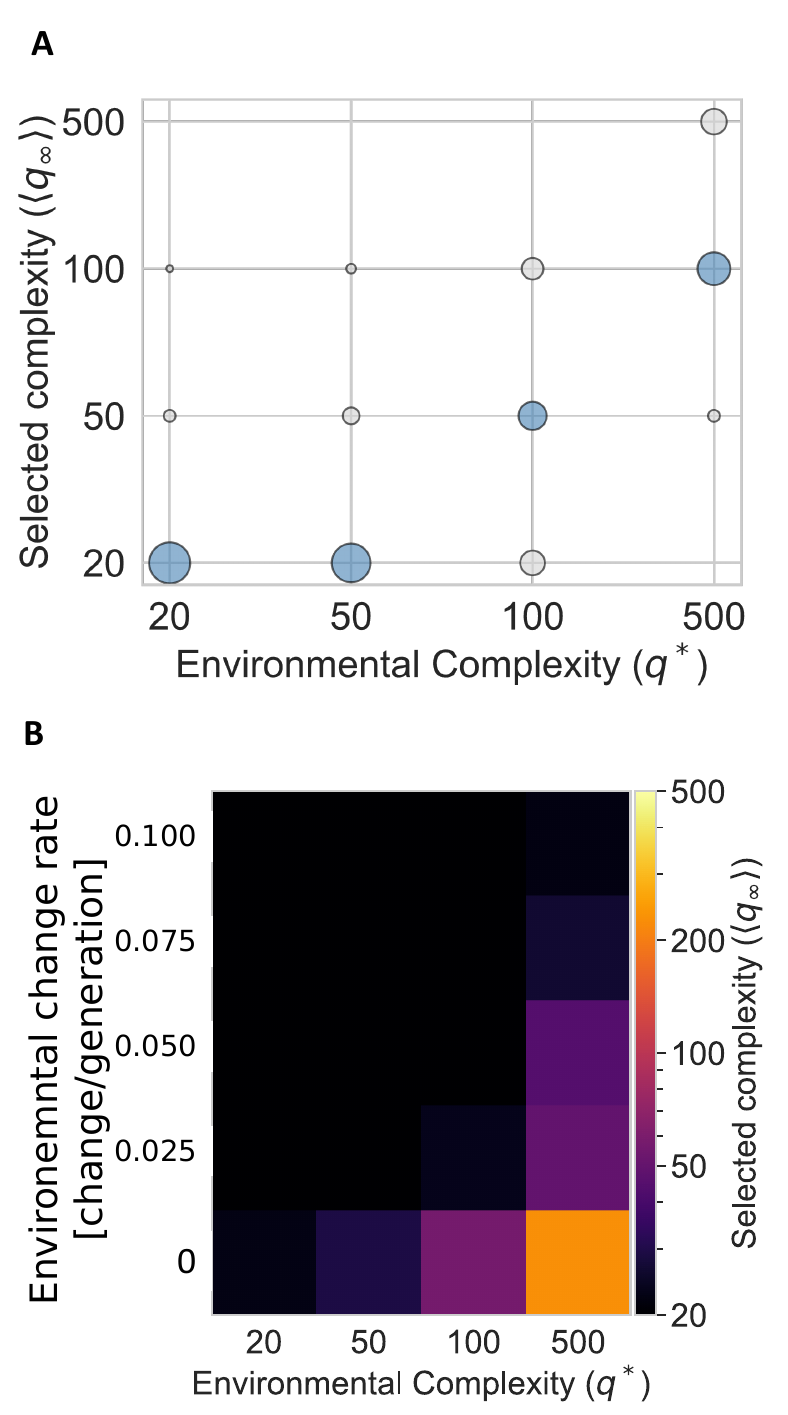}
\caption{\textbf{Selected complexity in stable and changing environments for neural networks}. 
\textbf{(A)} Circle sizes represent, for a given $q^*$ (x-axis), the fraction out of 100 simulations in which time-averaged maximum fitness was found in a given complexity class (y-axis), with highest fraction marked in blue.
\textbf{(B)} Selected class (average class at the end of selection) as a function of both environmental complexity (x-axis) and environmental change rate (here, the number of different environments in a simulation of length $T=1000$) (y-axis). 
Results are based on 100 simulation realizations. Population sizes of 2,000 per complexity class, $\sigma=1$, $\gamma=0.01$ and $T=1000$.  
}
\label{fig:NNs}
\end{figure}

Neural networks were chosen for their nature as universal functions approximators \cite{cybenko1989approximation}, and for their use as models of gene regulatory networks \cite{watson2014evolution}. While these functions do not have a one-to-one correspondence between environmental complexity and selected complexity as linear functions, they still show a trend of increasing selected complexity with increasing environmental complexity (fig \ref{fig:NNs}A). Changing environments has an effect of decreasing selected complexity (fig \ref{fig:NNs}B), as in the case of linear functions.

\section{\label{sec:app_pop_size}Effect of Population Size}
Function classes with more parameters are capable of more complex mappings, but they are also, by definition, larger. Most results in learning theory comparing function classes of different dimensionality assume that the entire continuum of possible parameterizations in each class is considered. In our case, however, we compare populations, which are finite-sized random samples from each class. 

If the optimal mapping $\phi^*$ contains $d_{opt}$ parameters and populations are of fixed size $n$ , then for any dimension $d > d_{opt}$ the probability of the population containing a member within some close distance to $\phi^*$ decreases with $d$ but increases with $n$.

\begin{figure}[t!]
\centering
\includegraphics[width=0.45\textwidth]{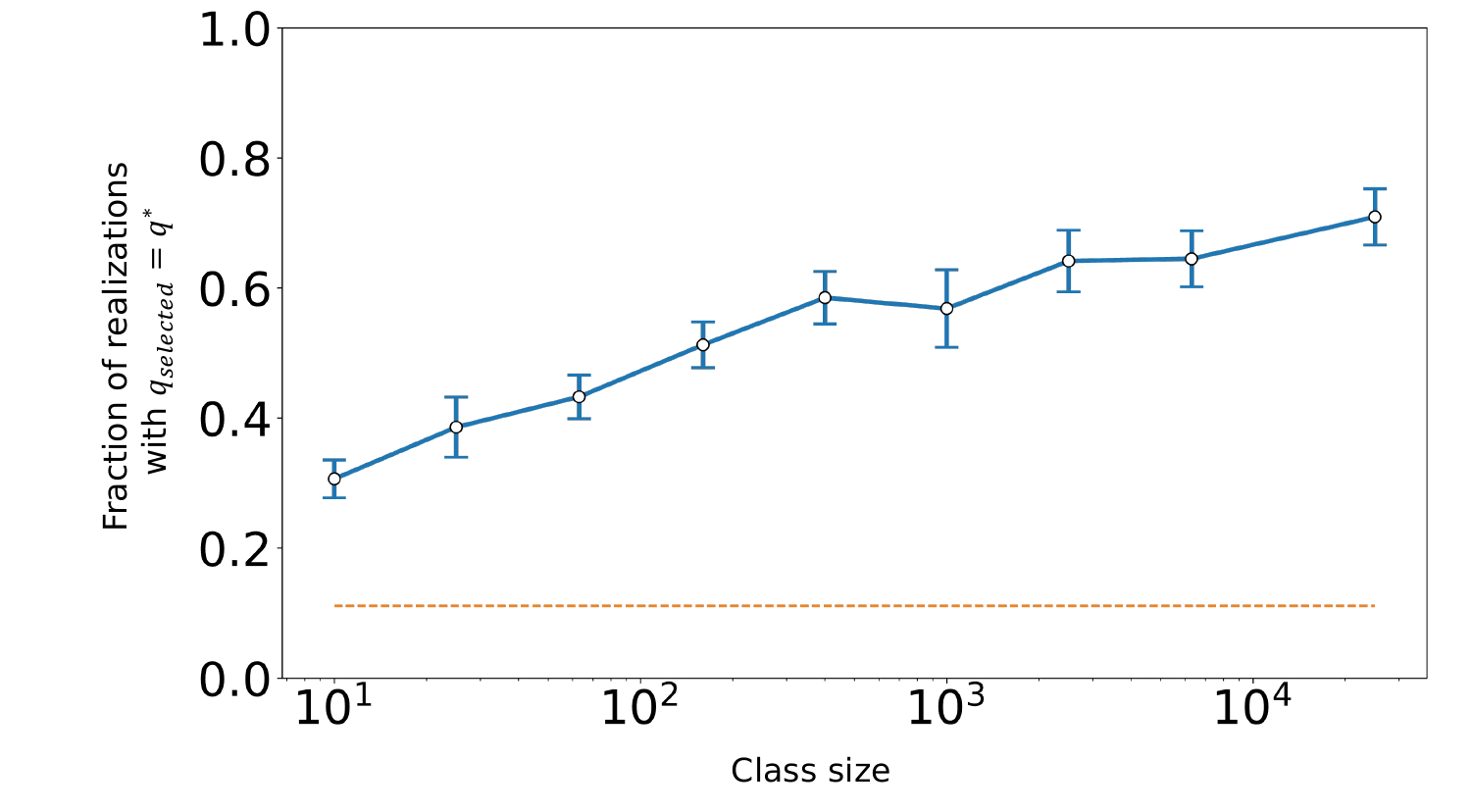}
\caption{\textbf{The effect of class size on the strength of implicit regularization.}
Simulations carried out over increasing population sizes (x-axis) of linear functions, result in increasingly larger fractions of realizations which end up with the dominant class at the end of selection equal to $q^*$.
Results from 100 simulation realization per value of $q^*$ in the range $[1, 9]$, with parameters $T=5,000$, $\gamma=0.05$ and $\sigma=1$.}
\label{fig:class_size}

\end{figure}

To illustrate this effect, we measured the correspondence between environmental complexity and selected complexity in linear functions, as described in the main text, but for varying population sizes. For each given population size, we ran 900 simulation realizations, 100 for each $q^*$ in the range $[1,9]$, and recorded the fraction of simulations where the selected complexity matched $q^*$ (fig \ref{fig:class_size}). As expected, we observed a trend of increasing "precision" of the selection process for increasing population size.

\section{\label{sec:app_init_cond}Effect of Initial Conditions}
\begin{figure*}[t!]
\centering
\includegraphics[width=0.9\textwidth]{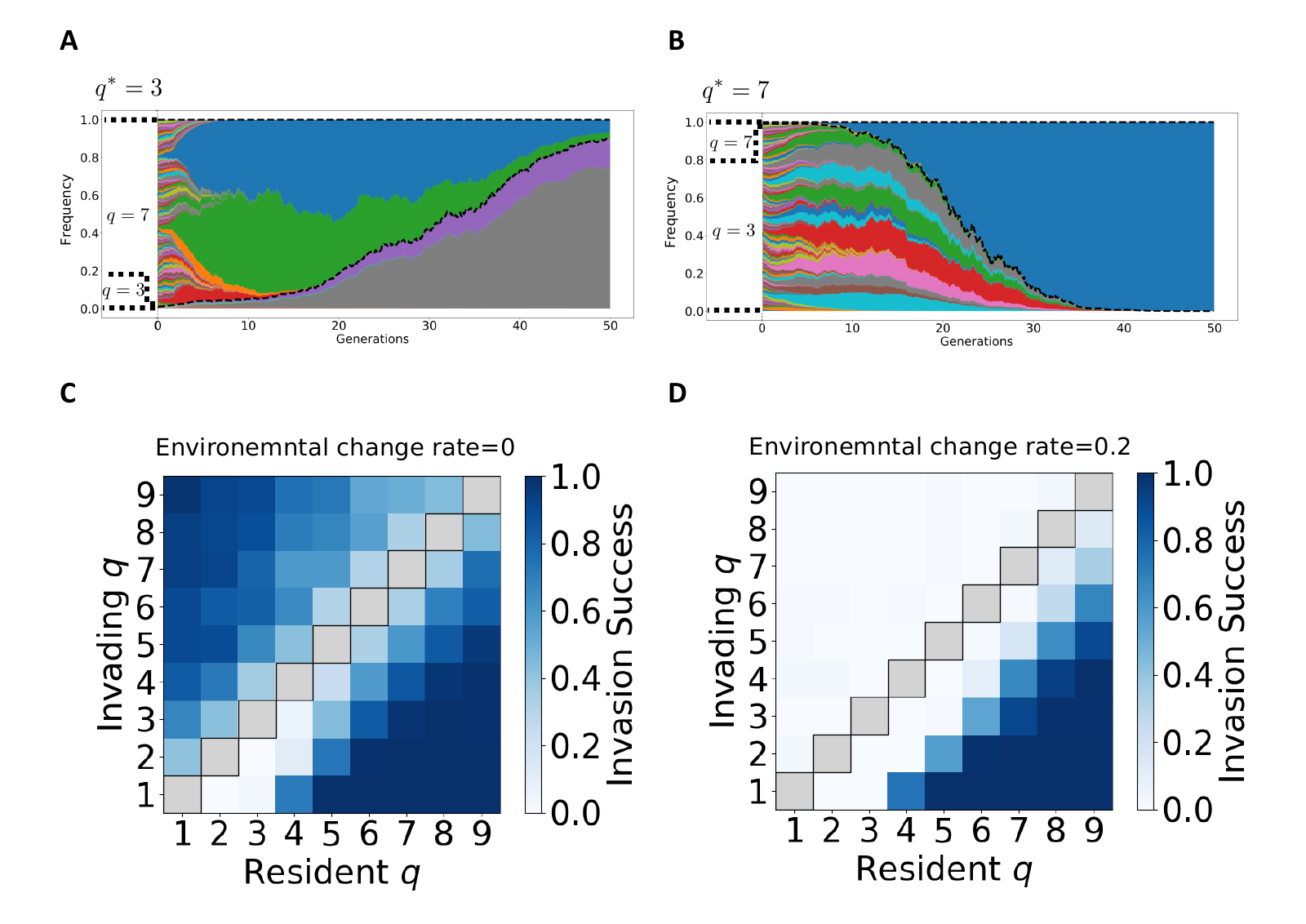}
\caption{\textbf{Invasion Simulations.}
\textbf{(A,B)} Resident populations whose complexity mismatches environmental complexity are typically successfully invaded by a small ($1\%$) population which matches the environmental complexity.\textbf{ (C)} Invasion success (fraction out of 200 invasion simulations, similar to (A, B), in which the invading population dominates at the end of selection) for all pairs of resident and invading complexity, where the environmental complexity, $q^*$, is always equal to the  invading complexity.\textbf{ (D)} Same as (C) but with changing environments at a rate of $0.2$ changes per generation.
Results in \textbf{(C,D)} from 200 simulation realizations per resident-invading pair. All results in \textbf{(A,B,C,D)} with population sizes of 2,000 per complexity class,  $T=1,000$, $\gamma=0.05$ and $\sigma=1$.}
\label{fig:invasion}

\end{figure*}

All results in the main text refer to uniform initial population composition, both within and between complexity classes. To demonstrate the robustness of the main findings to this assumption, we consider \textit{invasion experiments}, where a resident population of the "wrong" complexity (one which does not match environmental complexity) is seeded at $t_0$ with a small ($1\%$) subpopulation whose complexity is $q^*$ (equal to the underlying environmental complexity). Typical realizations of this setting results in the invading population dominating (fig \ref{fig:invasion}A-C). In the case of rapidly changing environments, however, simple invading populations are able to invade while complex invaders cannot (fig \ref{fig:invasion}D), as when the environment changes faster than the transient discussed in the main text (fig \ref{fig:changing_env} and accompanying text), a complex invading population, even if it contains the optimal type for any given environment, is unable to take hold before the environment changes.

\section{\label{sec:app_sim_details}Simulation Details}

\subsection{Function Classes}
All function classes considered consist of mappings from environmental cues $\vec{\xi}$ from an input space of dimension $n$ to phenotypes $\phi(\vec{\xi})$ from an output space of dimension $p$. Parameters, for all functions described, are drawn for each population member independently from $\mathcal{N}(0,1)$. Results in the main text refer to populations of linear maps. Simulations described in figs \ref{fig:polynomials}, and  \ref{fig:NNs}  were carried over populations of size 2,000 per complexity class, for $T=1000$ timesteps. Simulations described in fig \ref{fig:class_size} were carried over different class sizes (x-axis) and $T=5,000$ timesteps. All simulations were performed with $\gamma=0.05$ and $\sigma=1$.

\subsubsection{Linear Maps} 
These are functions of the form $\phi:\vec{\xi}\mapsto A \vec{\xi}$  where $A$ is an $n\times p$ matrix ($n=p=3$ in the simulations shown in the main text). Complexity $q\in [0, n\cdot p]$ is defined as the number of non-zero entries in the matrix, added first to the main diagonal and then to the off-diagonals in a predetermined order - first the super-diagonal starting at (1, 2), then the one starting at (1, 3) etc. (so for the $3\times 3$ case, $q=3$ maps are the class of diagonal matrices, $q=4$ maps have non-zeros in their diagonal and in their (1, 2) entry etc.).

\subsubsection{Polynomials}
Here we consider the mappings $\phi$ as polynomials parametrized by coefficients $\{a_i\}_{i=1}^q$ and defined as the maps $$\phi:\vec{\xi}\mapsto\left(\begin{array}{c}
\sum_{i=1}^{q}a_{i}\xi_{1}\\
\sum_{i=1}^{q}a_{i}\xi_{2}\\
\vdots\\
\sum_{i=1}^{q}a_{i}\xi_{p}
\end{array}\right)$$

Parameters $a_i$ for all functions are sampled i.i.d. from $\mathcal{N}(0, 1)$.

\subsubsection{One-Hidden-Layer Neural Networks}
These are defined as $$\phi:\vec{\xi}\mapsto W^{\left(2\right)}\sigma\left(W^{\left(1\right)}\vec{\xi}\right)$$
where $W^{\left(1\right)}\in\mathbb{R}^{q\times n}$, $W^{\left(2\right)}\in\mathbb{R}^{p\times q}$ and $\sigma(x) = \frac{1}{1+e^{-x}}$.

Parameters $W_{i,j}$ for all functions are sampled i.i.d. from $\mathcal{N}(0, 1)$.

\subsection{Simulations} 
Simulation dynamics consisted of applying the replicator equation (eq. \ref{eqn:replicator}) to populations of functions as described in the main text. All Simulations were carried over populations of size 2,000 per complexity class, for $T=1000$ timesteps with parameters $\gamma=0.05$ and $\sigma=1$ except where explicitly mentioned otherwise. All Python code required for replicating the results is available via GitHub at \href{https://github.com/nitzanlab/Overfitness}{https://github.com/nitzanlab/Overfitness}.

\section*{\label{sec:appendix}APPENDIX: DERIVATIONS}
\label{deriv}
\subsubsection{Class Fitness}
In a population evolving according to eq.  \ref{eqn:replicator} the frequency of any member $i$ in the next generation, denoted $x_{i}'$ is given by
$$x'_{i}=\frac{x_{i}f_{i}}{\sum_{j}x_{j}f_{j}}$$

The frequency $X_{k}$ of any class $C_{k}$ is thus 

$$X_{k}'=\sum_{i\in C_{k}}x_{i}'=\frac{\sum_{i\in C_{k}}x_{i}f_{i}}{\sum_{j}x_{j}f_{j}}=\frac{\sum_{i\in X_{k}}x_{i}f_{i}}{\sum_{l}\sum_{j\in C_{l}}x_{j}f_{j}}$$

If we define
$$F_{k}=\frac{\sum_{i\in C_{k}}x_{i}f_{i}}{X_{k}}=\frac{\sum_{i\in C_{k}}x_{i}f_{i}}{\sum_{i\in C_{k}}x_{i}}$$

we recover  eq.  \ref{eqn:replicator_class} 

$$X_{k}'=\frac{X_{k}F_{k}}{\sum_{l}X_{l}F_{l}}\propto X_{k}F_{k}$$

Thus $F_{k}$ plays an analogous role to individual member fitness$ f_{i}$, but at the class level. 

\subsubsection{Occam Factor}
For any $i\in C_{k}$ let $\tilde{x}_{i}=\frac{x_{i}}{X_{k}}$ (i.e. normalized to class frequency). We have 
$$f_{i}\frac{\tilde{x}_{i}}{\tilde{x}_{i}'}=f_{i}\frac{x_{i}}{x_{i}'}\frac{X_{k}'}{X_{k}}\overset{_{\left(1\right)}}{=}\sum_{j}x_{j}f_{j}\frac{X_{k}'}{X_{k}}\overset{_{\left(2\right)}}{=}\frac{\sum_{i\in C_{k}}x_{i}f_{i}}{\sum_{i\in C_{k}}x_{i}}=F_{k}$$

(1) Substituting $x'_{i}$ for $\frac{x_{i}f_{i}}{\sum_{j}x_{j}f_{j}}$ \\
(2) Substituting $X_{k}'$ for $\frac{\sum_{i\in X_{k}}x_{i}f_{i}}{\sum_{j}x_{j}f_{j}}$

Denoting $i_{k}^{*}\left(t\right)$ as the index of the member with highest fitness at time $t$ in class $k$ , we recover eq.  \ref{eqn:occam}.

\subsubsection{Convergence to Globally Optimal Member}
For any population evolving according to eq.  \ref{eqn:replicator} we have 
$$x_{i}\left(T\right)\propto x_{i}\left(0\right)\prod_{t=1}^{T}f_{i}\left(t\right)$$
and therefore 
$$\log x_{i}\left(T\right)=\log\left(x_{i}\left(0\right)\right)+\sum_{t=1}^{T}\log\left(f_{i}\left(t\right)\right)-Z$$
where $Z$ is a normalizing constant. For sufficiently large $T$, by the law of large numbers we have for any $i, j$
$$\log x_{i}\left(T\right)-\log x_{j}\left(T\right)=T\left(\left\langle \log f_{i}\right\rangle -\left\langle \log f_{j}\right\rangle \right)+o\left(1\right)$$
and thus, denoting $i^*$ as the index of the member with the largest mean log fitness, we have 
$$x_{i^{*}}\left(t\right)\xrightarrow[t\to\infty]{}1$$
and for the population's mean fitness $F$ , 
$$F\left(t\right)\xrightarrow[t\to\infty]{}f_{i^{*}}$$

\subsubsection{Fisher's Fundamental Theorem}
For a class $C_k$ evolving according to eq.  \ref{eqn:replicator} with fixed fitness over time we have 
\[
\begin{aligned}
F_{k}(t+1) 
&= \sum_{i \in C_k} \tilde{x}_i(t+1) f_i(t+1) \\
&= \frac{\sum_{i \in C_k} \tilde{x}_i(t) f_i(t) f_i(t+1)}{\sum_{j \in C_k} \tilde{x}_j(t) f_j(t)} \\
&= \frac{\sum_{i \in C_k} \tilde{x}_i(t) f_i^2}{\sum_{j \in C_k} \tilde{x}_j(t) f_j} \\
&= \frac{\sum_{i \in C_k} \tilde{x}_i(t) f_i^2}{F_k(t)}
\end{aligned}
\]
and therefore 

$$\Delta F_{k}\left(t\right)=F_{k}\left(t+1\right)-F_{k}\left(t\right)=\frac{\sum_{i\in C_{k}}\tilde{x}_{i}\left(t\right)f_{i}^{2}-F_{k}^{2}\left(t\right)}{F_{k}\left(t\right)}$$
The values $f_i$ are assumed to be sampled from a distribution $p_k$. At time $0$, when class frequencies $x_i$ are uniform, we have 

$$\left\langle \Delta F_{k}\left(0\right)\right\rangle _{p_{k}}=\frac{\left\langle f^{2}\right\rangle _{p_{k}}-\left\langle f\right\rangle _{p_{k}}^{2}}{\left\langle f\right\rangle _{p_{k}}}=\frac{Var\left[p_{k}\right]}{\left\langle p_{k}\right\rangle }$$

\bibliographystyle{unsrt}

\bibliography{paper}

\begin{thebibliography}{10}

\bibitem{kauffman_origins_2011}
Stuart~A. Kauffman.
\newblock {\em The origins of order: self-organization and selection in evolution}.
\newblock Oxford Univ. Pr, New York, nachdr. edition, 2011.

\bibitem{mcshea_complexity_1991}
Daniel~W. McShea.
\newblock Complexity and evolution: {What} everybody knows.
\newblock {\em Biology \& Philosophy}, 6(3):303--324, July 1991.

\bibitem{wolf2018physical}
Yuri~I Wolf, Mikhail~I Katsnelson, and Eugene~V Koonin.
\newblock Physical foundations of biological complexity.
\newblock {\em Proceedings of the National Academy of Sciences}, 115(37):E8678--E8687, 2018.

\bibitem{szathmary1995major}
E{\"o}rs Szathm{\'a}ry and John~Maynard Smith.
\newblock The major evolutionary transitions.
\newblock {\em Nature}, 374(6519):227--232, 1995.

\bibitem{adami2000evolution}
Christoph Adami, Charles Ofria, and Travis~C Collier.
\newblock Evolution of biological complexity.
\newblock {\em Proceedings of the National Academy of Sciences}, 97(9):4463--4468, 2000.

\bibitem{darwin1859}
Charles Darwin.
\newblock {\em On the Origin of Species by Means of Natural Selection}.
\newblock Murray, London, 1859.
\newblock or the Preservation of Favored Races in the Struggle for Life.

\bibitem{spencer_principles_2020}
Herbert Spencer.
\newblock {\em The {Principles} of {Biology}: {Volume} 1}.
\newblock Outlook Verlag, 2020.

\bibitem{lenski_evolutionary_2003}
Richard~E. Lenski, Charles Ofria, Robert~T. Pennock, and Christoph Adami.
\newblock The evolutionary origin of complex features.
\newblock {\em Nature}, 423(6936):139--144, May 2003.

\bibitem{carroll_chance_2001}
Sean~B. Carroll.
\newblock Chance and necessity: the evolution of morphological complexity and diversity.
\newblock {\em Nature}, 409(6823):1102--1109, February 2001.

\bibitem{keeling_simplicity_2004}
Patrick~J. Keeling and Claudio~H. Slamovits.
\newblock Simplicity and {Complexity} of {Microsporidian} {Genomes}.
\newblock {\em Eukaryotic Cell}, 3(6):1363--1369, December 2004.

\bibitem{veit2025evolution}
Walter Veit, Samuel~JL Gascoigne, and Roberto Salguero-G{\'o}mez.
\newblock Evolution, complexity, and life history theory.
\newblock {\em Biological Theory}, pages 1--10, 2025.

\bibitem{wagner_energy_2005}
Andreas Wagner.
\newblock Energy {Constraints} on the {Evolution} of {Gene} {Expression}.
\newblock {\em Molecular Biology and Evolution}, 22(6):1365--1374, June 2005.

\bibitem{lynch_bioenergetic_2024}
Michael Lynch.
\newblock The bioenergetic cost of building a metazoan.
\newblock {\em Proceedings of the National Academy of Sciences}, 121(46):e2414742121, November 2024.

\bibitem{muller_genetic_1932}
H.~J. Muller.
\newblock Some {Genetic} {Aspects} of {Sex}.
\newblock {\em The American Naturalist}, 66(703):118--138, March 1932.
\newblock Publisher: The University of Chicago Press.

\bibitem{moran_accelerated_1996}
N~A Moran.
\newblock Accelerated evolution and {Muller}'s rachet in endosymbiotic bacteria.
\newblock {\em Proceedings of the National Academy of Sciences}, 93(7):2873--2878, April 1996.

\bibitem{maughan2007roles}
Heather Maughan, Joanna Masel, C~William Birky~Jr, and Wayne~L Nicholson.
\newblock The roles of mutation accumulation and selection in loss of sporulation in experimental populations of bacillus subtilis.
\newblock {\em Genetics}, 177(2):937--948, 2007.

\bibitem{heylighen_growth_1999}
Francis Heylighen.
\newblock The growth of structural and functional complexity during evolution.
\newblock {\em The evolution of complexity}, 8:17--44, 1999.
\newblock Publisher: Kluwer Dordrecht.

\bibitem{orr_adaptation_2000}
H.~Allen Orr.
\newblock {ADAPTATION} {AND} {THE} {COST} {OF} {COMPLEXITY}.
\newblock {\em Evolution}, 54(1):13--20, February 2000.

\bibitem{wagner2008pleiotropic}
G{\"u}nter~P Wagner, Jane~P Kenney-Hunt, Mihaela Pavlicev, Joel~R Peck, David Waxman, and James~M Cheverud.
\newblock Pleiotropic scaling of gene effects and the ‘cost of complexity’.
\newblock {\em Nature}, 452(7186):470--472, 2008.

\bibitem{adami_what_2002}
Christoph Adami.
\newblock What is complexity?
\newblock {\em BioEssays}, 24(12):1085--1094, December 2002.

\bibitem{hinegardner1983biological}
Ralph Hinegardner and Joseph Engelberg.
\newblock Biological complexity.
\newblock {\em Journal of Theoretical Biology}, 104(1):7--20, 1983.

\bibitem{grunwald_minimum_2007}
Peter~D. Grünwald.
\newblock {\em The minimum description length principle}.
\newblock Adaptive computation and machine learning. MIT Press, Cambridge, Mass, 2007.

\bibitem{simon_architecture_1962}
HERBERT~A SIMON.
\newblock {THE} {ARCHITECTURE} {OF} {COMPLEXITY}.
\newblock {\em PROCEEDINGS OF THE AMERICAN PHILOSOPHICAL SOCIETY}, 106(6), 1962.

\bibitem{dingle_inputoutput_2018}
Kamaludin Dingle, Chico~Q. Camargo, and Ard~A. Louis.
\newblock Input–output maps are strongly biased towards simple outputs.
\newblock {\em Nature Communications}, 9(1):761, February 2018.
\newblock Publisher: Nature Publishing Group.

\bibitem{godfrey-smith_complexity_1998}
Peter Godfrey-Smith.
\newblock {\em Complexity and the {Function} of {Mind} in {Nature}}.
\newblock Cambridge University Press, 1998.

\bibitem{bird1995gene}
Adrian~P Bird.
\newblock Gene number, noise reduction and biological complexity.
\newblock {\em Trends in Genetics}, 11(3):94--100, 1995.

\bibitem{brem2005landscape}
Rachel~B Brem and Leonid Kruglyak.
\newblock The landscape of genetic complexity across 5,700 gene expression traits in yeast.
\newblock {\em Proceedings of the National Academy of Sciences}, 102(5):1572--1577, 2005.

\bibitem{van2019biological}
Marc~HV Van~Regenmortel and Marc~HV Van~Regenmortel.
\newblock Biological complexity emerges from the ashes of genetic reductionism.
\newblock {\em HIV/AIDS: Immunochemistry, Reductionism and Vaccine Design: A Review of 20 Years of Research}, pages 79--86, 2019.

\bibitem{mcshea1996perspective}
Daniel~W McShea.
\newblock Perspective metazoan complexity and evolution: is there a trend?
\newblock {\em Evolution}, 50(2):477--492, 1996.

\bibitem{valentine1994morphological}
James~W Valentine, Allen~G Collins, and C~Porter Meyer.
\newblock Morphological complexity increase in metazoans.
\newblock {\em Paleobiology}, 20(2):131--142, 1994.

\bibitem{nowak_evolutionary_2006}
Martin~A. Nowak.
\newblock {\em Evolutionary dynamics: exploring the equations of life}.
\newblock The Belknap Press of Harvard University Press, Cambridge, Massachusetts London, England, 2006.

\bibitem{harper_replicator_2010}
Marc Harper.
\newblock The {Replicator} {Equation} as an {Inference} {Dynamic}, May 2010.
\newblock arXiv:0911.1763 [math].

\bibitem{watson_how_2016}
Richard~A Watson and Eörs Szathmáry.
\newblock How can evolution learn?
\newblock {\em Trends in ecology \& evolution}, 31(2):147--157, 2016.
\newblock Publisher: Elsevier.

\bibitem{kouvaris_how_2017}
Kostas Kouvaris, Jeff Clune, Loizos Kounios, Markus Brede, and Richard~A Watson.
\newblock How evolution learns to generalise: {Using} the principles of learning theory to understand the evolution of developmental organisation.
\newblock {\em PLoS computational biology}, 13(4):e1005358, 2017.
\newblock Publisher: Public Library of Science San Francisco, CA USA.

\bibitem{czegel_bayes_2022}
Dániel Czégel, Hamza Giaffar, Joshua~B. Tenenbaum, and Eörs Szathmáry.
\newblock Bayes and {Darwin}: {How} replicator populations implement {Bayesian} computations.
\newblock {\em BioEssays}, 44(4):2100255, April 2022.

\bibitem{campbell_universal_2016}
John~O Campbell.
\newblock Universal {Darwinism} as a process of {Bayesian} inference.
\newblock {\em Frontiers in systems neuroscience}, 10:49, 2016.
\newblock Publisher: Frontiers Media SA.

\bibitem{watson2014evolution}
Richard~A Watson, G{\"u}nter~P Wagner, Mihaela Pavlicev, Daniel~M Weinreich, and Rob Mills.
\newblock The evolution of phenotypic correlations and “developmental memory”.
\newblock {\em Evolution}, 68(4):1124--1138, 2014.

\bibitem{akaike_new_1974}
H.~Akaike.
\newblock A new look at the statistical model identification.
\newblock {\em IEEE Transactions on Automatic Control}, 19(6):716--723, December 1974.

\bibitem{burnham_model_2004}
Kenneth~P. Burnham and David~R. Anderson, editors.
\newblock {\em Model {Selection} and {Multimodel} {Inference}}.
\newblock Springer New York, New York, NY, 2004.

\bibitem{vapnik_nature_1995}
Vladimir~N. Vapnik.
\newblock {\em The {Nature} of {Statistical} {Learning} {Theory}}.
\newblock Springer New York, New York, NY, 1995.

\bibitem{bishop_pattern_2006}
Christopher~M Bishop and Nasser~M Nasrabadi.
\newblock {\em Pattern recognition and machine learning}, volume~4.
\newblock Springer, 2006.

\bibitem{mackay_information_2003}
David~JC MacKay.
\newblock {\em Information theory, inference and learning algorithms}.
\newblock Cambridge university press, 2003.

\bibitem{wolpert_rigorous_1992}
David~H Wolpert.
\newblock A rigorous investigation of “evidence” and “{Occam} factors” in {Bayesian} reasoning.
\newblock {\em The Sante Fe Institute}, 1660, 1992.

\bibitem{malnic1999combinatorial}
Bettina Malnic, Junzo Hirono, Takaaki Sato, and Linda~B Buck.
\newblock Combinatorial receptor codes for odors.
\newblock {\em Cell}, 96(5):713--723, 1999.

\bibitem{xue_environment--phenotype_2019}
BingKan Xue, Pablo Sartori, and Stanislas Leibler.
\newblock Environment-to-phenotype mapping and adaptation strategies in varying environments.
\newblock {\em Proceedings of the National Academy of Sciences}, 116(28):13847--13855, July 2019.

\bibitem{fisher1999genetical}
Ronald~Aylmer Fisher.
\newblock {\em The genetical theory of natural selection: a complete variorum edition}.
\newblock Oxford University Press, 1999.

\bibitem{burger1995evolution}
Reinhard B{\"u}rger and Michael Lynch.
\newblock Evolution and extinction in a changing environment: a quantitative-genetic analysis.
\newblock {\em Evolution}, 49(1):151--163, 1995.

\bibitem{chevin2010adaptation}
Luis-Miguel Chevin, Russell Lande, and Georgina~M Mace.
\newblock Adaptation, plasticity, and extinction in a changing environment: towards a predictive theory.
\newblock {\em PLoS biology}, 8(4):e1000357, 2010.

\bibitem{mustonen_fitness_2009}
Ville Mustonen and Michael Lässig.
\newblock From fitness landscapes to seascapes: non-equilibrium dynamics of selection and adaptation.
\newblock {\em Trends in Genetics}, 25(3):111--119, March 2009.

\bibitem{dridi2016environmental}
Slimane Dridi and Laurent Lehmann.
\newblock Environmental complexity favors the evolution of learning.
\newblock {\em Behavioral Ecology}, 27(3):842--850, 2016.

\bibitem{godfrey2013environmental}
Peter Godfrey-Smith.
\newblock Environmental complexity and the evolution of cognition.
\newblock In {\em The evolution of intelligence}, pages 223--249. Psychology Press, 2013.

\bibitem{dawkins1979arms}
Richard Dawkins and John~Richard Krebs.
\newblock Arms races between and within species.
\newblock {\em Proceedings of the Royal Society of London. Series B. Biological Sciences}, 205(1161):489--511, 1979.

\bibitem{bergman1994evolution}
Aviv Bergman and Marcus~W Feldman.
\newblock On the evolution of learning.
\newblock {\em Theor. Pop. Biol}, 1994.

\bibitem{bergland_genomic_2014}
Alan~O. Bergland, Emily~L. Behrman, Katherine~R. O'Brien, Paul~S. Schmidt, and Dmitri~A. Petrov.
\newblock Genomic {Evidence} of {Rapid} and {Stable} {Adaptive} {Oscillations} over {Seasonal} {Time} {Scales} in {Drosophila}.
\newblock {\em PLoS Genetics}, 10(11):e1004775, November 2014.

\bibitem{hastie2009elements}
Trevor Hastie, Robert Tibshirani, Jerome~H Friedman, and Jerome~H Friedman.
\newblock {\em The elements of statistical learning: data mining, inference, and prediction}, volume~2.
\newblock Springer, 2009.

\bibitem{cybenko1989approximation}
George Cybenko.
\newblock Approximation by superpositions of a sigmoidal function.
\newblock {\em Mathematics of control, signals and systems}, 2(4):303--314, 1989.

\end{thebibliography}
\end{document}